\documentclass[aip,reprint,amsmath,amssymb,superscriptaddress,floatfix,citeautoscript]{revtex4-2}
\usepackage{graphicx}
\usepackage{color}
\usepackage{bm}
\usepackage{braket}
\usepackage{tabularx}
\usepackage{multirow}
\usepackage[colorlinks,linkcolor=blue,citecolor=blue,urlcolor=blue]{hyperref}

\begin{document}

\title{Rational Design of Efficient Defect-Based Quantum Emitters}

\author{Mark E. Turiansky}
\email{mturiansky@ucsb.edu}
\affiliation{Materials Department, University of California, Santa Barbara, CA 93106-5050, U.S.A.}

\author{Kamyar Parto}
\affiliation{Department of Electrical and Computer Engineering, University of California, Santa Barbara, CA 93106-5080, U.S.A.}

\author{Galan Moody}
\affiliation{Department of Electrical and Computer Engineering, University of California, Santa Barbara, CA 93106-5080, U.S.A.}
 
\author{Chris G. Van de Walle}
\email{vandewalle@mrl.ucsb.edu}
\affiliation{Materials Department, University of California, Santa Barbara, CA 93106-5050, U.S.A.}

\date{\today}

\begin{abstract}
    Single-photon emitters are an essential component of quantum networks, and
    defects or impurities in semiconductors are a promising platform to realize such quantum emitters.
    Here we present a model that encapsulates the essential physics of coupling to phonons, which governs the behavior of real single-photon emitters,
    and critically evaluate several approximations that are commonly utilized.
    Emission in the telecom wavelength range is highly desirable, but our model shows that
    nonradiative processes are greatly enhanced at these low photon energies, leading to a decrease in efficiency.
    Our results suggest that reducing the phonon frequency is a fruitful avenue to enhance the efficiency.
\end{abstract}

\maketitle

\section{Introduction}

The generation of single photons is essential to power the second quantum revolution and to realize the promise of the quantum internet~\cite{obrien_photonic_2009,northup_quantum_2014-1,wehner_quantum_2018}.
Photons are natural carriers of quantum information.
They have very weak interactions with the environment and, thanks to decades of development in fiber-optic technologies, can be transmitted over long distances with minimal loss.
An ideal single-photon emitter acts as a ``photon gun'', producing single photons on demand.
Three properties should be optimized~\cite{esmann_solid-state_2023}:
(i) brightness, as quantified through the intensity and efficiency,
(ii) photon purity, as quantified through the photon autocorrelation function measured in a Hanbury Brown-Twiss experiment,
and (iii) indistinguishability, as quantified through the ability of photons to interfere in a Hong-Ou-Mandel experiment.
In addition, if long-range transmission is required,
the photon energies should lie within telecom wavelengths, 1260~nm to~1675 nm (between 0.74 and 0.98~eV), to take advantage of fiber optics~\cite{agrawal_fiber-optic_2010}.

Single photons in a well-defined quantum state can be produced in a variety of ways;
point defects embedded in a semiconductor or insulator are a particularly promising platform.
(We use the term ``point defects'' to refer to both native defects, which are intrinsic to the lattice, as well as extrinsic impurities or a complex of the two.)
Point defects can be used as qubits, quantum memories, or single-photon emitters, all essential components of quantum networks, and have been demonstrated to operate even at room temperature~\cite{dreyer_first-principles_2018,aharonovich_solid-state_2016,borregaard_quantum_2019,bassett_quantum_2019-1}.

Long-range networking using defect-based single-photon emitters has been demonstrated~\cite{aharonovich_solid-state_2016,bhaskar_experimental_2020,bernien_heralded_2013},
mainly relying on the nitrogen-vacancy (NV) center in diamond~\cite{gali_ab_2019,janitz_cavity_2020}.
However it is known that the optical interface of the NV center is not ideal.
In particular, less than 3\% of the emitted photons are in the zero-phonon line (ZPL)---in other words, useful for quantum information---due to coupling to phonons.
This is a result of the interaction between the electronic states of the defect with the diamond host lattice, referred to as electron-phonon coupling.
Electron-phonon coupling also broadens the ZPL, leading to dephasing which reduces indistinguishability.
Alternatives to the NV center that have weaker electron-phonon coupling and stronger emission into the ZPL, such as the silicon-vacancy (SiV) center, are being pursued~\cite{bhaskar_experimental_2020,janitz_cavity_2020,aharonovich_solid-state_2016,bassett_quantum_2019-1}.

In addition to its impact on the ZPL, electron-phonon coupling has a second effect, namely the introduction of nonradiative decay.
After a photon is absorbed at a quantum defect, the system is in an excited state.
Ideally, it will decay radiatively by emitting a photon, but it may also decay through some other mechanism; nonradiative decay mediated by electron-phonon coupling can lead to an alternative recombination channel that may dominate over the radiative process.

The NV center produces photons in the visible spectrum; for many applications, emission at lower energies is desirable.
As noted above, telecom-wavelength emitters are desirable for long-range networking.
In addition, cavity coupling is often required to enhance brightness~\cite{lee_integrated_2020}, and high-quality cavities are easier to fabricate at longer wavelengths.
Numerous luminescent centers in diamond have been observed and characterized~\cite{zaitsev_optical_2001}, yet centers that produce telecom-wavelength photons are rare.
More generally, reports of longer-wavelength single-photon emitters in any material are scarce, which is surprising given the ubiquity of defects.
In diamond, it has been previously suggested that this may be due to strong electron-phonon coupling, which causes nonradiative processes to dominate~\cite{rogers_how_2010}.
To our knowledge, no rigorous investigation of this proposal or its extension to other materials exists.

Here we develop a model that captures the essential role that electron-phonon coupling plays in point-defect-based single-photon emitters.
Our results show that nonradiative processes dominate at smaller transition energies and indicate that obtaining a high-efficiency single-photon emitter at longer wavelengths is indeed difficult.
However, our model allows us to suggest productive avenues for improving the efficiency; one example is by reducing the average phonon frequency.
We also assess several approximations commonly employed in the literature surrounding the evaluation of emission rates.
The perspective provided by our model sheds new light on results in the existing literature.

\section{Basic Properties}
\begin{figure}[!htb]
    \centering
    \includegraphics[width=\columnwidth,height=0.5\textheight,keepaspectratio]{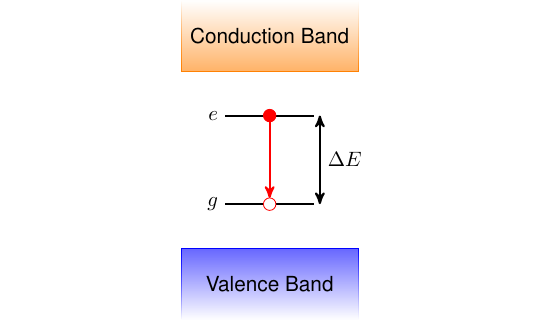}
    \caption{\label{fig:tls}
        Energy level diagram for a transition between intra-defect states.
        The ground $g$ and excited $e$ states are labeled.
        Orange corresponds to the conduction band and blue to the valence band.
    }
\end{figure}

We consider point defects embedded in a semiconductor or insulator.
Defects often give rise to states that lie well within the band gap~\cite{dreyer_first-principles_2018,aharonovich_solid-state_2016}.
For a given charge state of the defect, these states are occupied with electrons, and in some cases an electron can be excited from an occupied to an unoccupied state of the defect, thus defining the ground state ($g$) and excited state ($e$) of a two-level system, separated by an energy $\Delta E$ [Fig.~\ref{fig:tls}(a)].

The transition energy $\Delta E$ of the emitter is an important factor in the long-range transmission efficiency.
A single-photon emitter that produces photons within the telecom wavelength range of 1260--1675~nm ($\Delta E$ between 0.74 and 0.98~eV) could take advantage of fiber optics~\cite{agrawal_fiber-optic_2010}, with loss values below 0.2~dB/km.
Free-space communication is another option~\cite{kaushal_optical_2017}.
Several low-loss atmospheric windows exist;
one commonly utilized window covers wavelengths of 780--850~nm ($\Delta E$ between 1.46 and 1.59~eV).

An important issue for an ideal single-photon emitter is stability.
For an emitter based on point defects, one important form of stability is charge-state stability:
the ideal emitter avoids exchanging electrons with the valence or conduction bands.
Processes that result in an exchange of electrons with the bulk bands are referred to as charge dynamics.
Charge dynamics can occur nonradiatively (through thermal ionization or capture processes) or radiatively (through the absorption or emission of a photon).
To avoid charge dynamics, the deep defect states should be far from the band edges, for both the ground and excited states.
Such a system can be operated at elevated temperatures, a feature highly desirable for quantum applications.

It is worth noting that a single-photon emitter based on point defects can also be realized with involvement of a carrier in the valence or conduction band, bound to the defect in an excitonic state~\cite{haynes_experimental_1960,linhart_localized_2019,zhang_optically_2020},
i.e., the excited state involves an electron or hole in a hydrogenic wavefunction.
Alternatively, a bulk exciton (in which the electron and hole reside in band states) can be bound to a point defect~\cite{haynes_experimental_1960,abraham_shallow_2019,fu_ultrafast_2008,linpeng_coherence_2018}.
Such systems are inherently susceptible to charge dynamics since they arise from interactions with the bulk states of the host material.
The binding energy will determine the overall stability, but such emitters typically need be operated at low temperature.

Spin dynamics occur when the total spin of the defect changes, which is another form of stability that needs to be addressed.
Due to the weak nature of the interactions involved in changing the spin of the defect, spin dynamics can lead to the defect being in a dark, non-emissive state for long periods of time.
Thus spin dynamics can be detrimental to the efficiency of a single-photon emitter and should be avoided, for instance by selecting a defect with a level structure that precludes changes of spin.
We note that spin dynamics is not always harmful; there can be advantages to having a single-photon emitter with a ground-state spin that can be manipulated through spin dynamics.
Indeed, this is an essential feature of the NV center: a transition from the triplet to the singlet manifold enables optical manipulation of the ground-state spin.
In this paper, we focus on designing a highly efficient emitter, and for that purpose we will assume spin dynamics are negligible.

Given the above definitions, we can address two of the three parameters for an optimal single-photon emitter, namely purity and indistinguishability~\cite{esmann_solid-state_2023}.
Photon purity is evaluated in a Hanbury Brown-Twiss experiment, in which the photon autocorrelation function $g^{(2)} (\tau)$ is measured as a function of the time delay $\tau$ between the two light beams~\cite{fishman_photon-emission-correlation_2023}.
At short time delays, a single-photon emitter exhibits an ``antibunching dip'' with $1 - g^{(2)}(0)$ quantifying the photon purity, which should be unity for a perfect single-photon emitter.
Our idealized defect with only two levels will exhibit unity purity, assuming a single defect can be isolated.

Indistinguishability is the probability of two-photon interference in a Hong-Ou-Mandel experiment~\cite{grange_cavity-funneled_2015}.
Indistinguishability can be affected by spectral diffusion, which causes the transition energy to vary due to fluctuating charges in the vicinity of the emitter.
Resonant or quasi-resonant optical pumping is usually effective for reducing or eliminating spectral diffusion.
Here we assume our model defect to be an idealized two-level emitter that has a fixed transition energy, and thus unity indistinguishability in the absence of pure dephasing mechanisms.
Pure dephasing occurs when the quantum state picks up a phase from coupling to the local environment.
A dominant source of pure dephasing in solid-state systems is elastic scattering with acoustic phonons~\cite{muljarov_dephasing_2004,jahnke_electronphonon_2015}, which becomes negligible at low temperatures.
Throughout this work, we will assume low temperature, allowing us to assume unity indistinguishability for our model system.

The remaining parameter, brightness, is the main focus of this paper and is addressed in Sec.~\ref{sec:model}.

\section{Model}
\label{sec:model}

\subsection{Radiative Properties}
The rate at which a transition from the excited state to the ground state produces photons via the electric-dipole interaction is given by~\cite{stoneham_theory_1975,razinkovas_photoionization_2021-1}
\begin{equation}
    \label{eq:rad0}
    \Gamma_{\rm R}^{(0)} = {\left( \frac{\mathcal{E}_{\rm eff}}{\mathcal{E}_0} \right)}^2 \frac{n_r \mu^2 (\Delta E)^3}{3 \pi \epsilon_0 c^3 \hbar^4} \;,
\end{equation}
where $n_r$ is the index of refraction, which we will take to be 2.4, the value for diamond~\cite{levinshtein_handbook_1996}, which is also close to the value for boron nitride, silicon nitride, silicon carbide, and other insulator host materials.
$\mu$ is the transition dipole moment, which defines the strength of the transition.
Local-field effects describe the fact that the electric field at the defect may be different from that in bulk, due to scattering of light at the defect~\cite{stoneham_theory_1975}.
This is captured in the prefactor $\mathcal{E}_{\rm eff} / \mathcal{E}_0$, which is the ratio of the effective electric field at the defect to the bulk value.
Various models~\cite{smith_v_1972,stoneham_theory_1975} have been proposed to estimate this ratio; they produce a value larger than---but close to---one.
We will set $\mathcal{E}_{\rm eff} / \mathcal{E}_0$=1, a common assumption in the literature~\cite{razinkovas_photoionization_2021-1}.

\begin{figure}[!htb]
    \centering
    \includegraphics[width=\columnwidth,height=0.5\textheight,keepaspectratio]{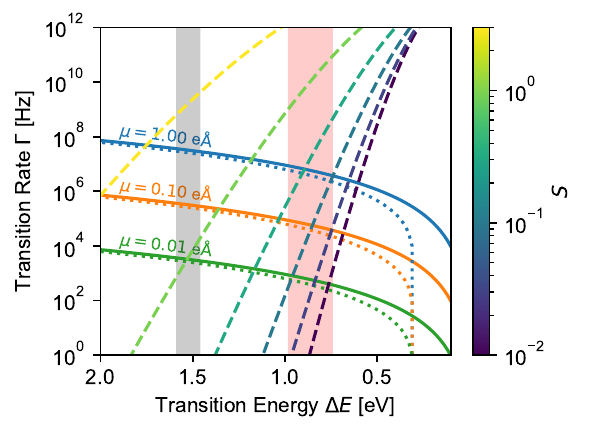}
    \caption{\label{fig:rate}
        The radiative rate without phonons $\Gamma_{\rm R}^{(0)}$ (solid) and with phonons $\Gamma_{\rm R}$ (dotted) for a transition dipole moment $\mu = 1$~e{\AA} (blue), 0.1~e{\AA} (orange), and 0.01~e{\AA} (green).
        The nonradiative rate $\Gamma_{\rm NR}$ is shown by the dashed lines for different values of the Huang-Rhys factor $S$ given by the color bar.
        For $\Gamma_{\rm R}$, the Huang-Rhys factor is assumed to be $S = 3$.
        Energies that fall within the range of telecom wavelengths are shaded in pink, and energies that fall within the free-space communication window are shaded in grey.
    }
\end{figure}

The transition dipole moment $\mu$ plays a key role in determining the overall radiative rate.
The lower limit on $\mu$ (and therefore $\Gamma_{\rm R}^{(0)}$) is zero, for a forbidden transition where the dipole moment is zero by symmetry.
An upper limit on $\mu$ can be estimated by introducing the concept of the oscillator strength $f$, given by
\begin{equation}
    \label{eq:osc_str}
    f = \frac{2 m_e (\Delta E)}{e^2 \hbar^2} \mu^2 \;.
\end{equation}
The oscillator strengths for transitions to all possible final states for a given initial state in the system must sum up to one (the Thomas-Kuhn sum rule~\cite{stoneham_theory_1975}).
If we assume a single transition dominates the sum, then $f \approx 1$ and for $\Delta E = 1$~eV, we find an upper limit on $\mu$ of $\approx$1.95~e{\AA}.
In practice, some of the largest transition dipole moments are found for transitions between orbitals of $s$ and $p$ character.
Such orbitals are common in $sp$-bonded covalent materials and have a transition dipole moment on the order of 1~e{\AA}.
For example, the transition dipole moment for the carbon dimer in hexagonal boron nitride (h-BN) is 1.06~e{\AA}~\cite{mackoit-sinkeviciene_carbon_2019}.
The blue line in Fig.~\ref{fig:rate} depicts Eq.~(\ref{eq:rad0}) evaluated for $\mu = 1$~e{\AA} as a function of $\Delta E$.

When symmetry or conservation rules lead to a forbidden electric-dipole transition, magnetic-dipole transitions may be observed, as is commonly the case for rare-earth impurities~\cite{dodson_magnetic_2012,stevenson_erbium-implanted_2022}.
While our focus is on electric-dipole transitions here, magnetic-dipole transitions have a similar cubic dependence on the energy~\cite{dodson_magnetic_2012}, and many of our arguments can be applied to such transitions with the magnitude of the radiative rate rescaled.

If this electronic system were isolated, its description would be complete.
However, the presence of the semiconducting or insulating host lattice provides a phonon bath for the electronic states to couple to.
Electron-phonon coupling has two main effects on our system:
(i) it gives rise to the phonon sideband in the luminescence spectrum, and (ii) it provides nonradiative pathways for energy to be dissipated.

\subsection{Electron-Phonon Coupling}

In a three-dimensional solid with $N$ atoms, there are 3$N$ vibrational modes to couple to.
Here we will study the interaction with a single, dispersionless phonon mode with energy $\hbar\Omega$.
A single-mode approximation has been effectively used as the basis of first-principles calculations of luminescence lineshapes~\cite{alkauskas_first-principles_2012} and radiative~\cite{dreyer_radiative_2020} and nonradiative capture rates~\cite{alkauskas_first-principles_2014,turiansky_nonrad_2021};
such calculations have achieved both a qualitative and quantitative description of experimental observations~\cite{alkauskas_first-principles_2012,wickramaratne_comment_2018,shi_comparative_2015}.
A single-mode approximation can thus be considered a realistic approximation to the electron-phonon coupling problem.
The single mode in this approximation is known as the accepting mode~\cite{stoneham_non-radiative_1981,stoneham_non-radiative_1978}.
It is a local vibrational mode that is not necessarily a normal mode of the system and is dominated by the motion of the defect and its nearest neighbor atoms.
Such a mode is thus highly sensitive to the chemical nature of these atoms.
Unless otherwise stated, we will assume a phonon energy of $\hbar\Omega = 100$~meV, which is common for the first-row elements that are often present in  materials that are used as hosts for quantum defects.

\subsubsection{Phonon Sideband}
\label{sssec:PSB}

\begin{figure}[!htb]
    \centering
    \includegraphics[width=\columnwidth,height=0.5\textheight,keepaspectratio]{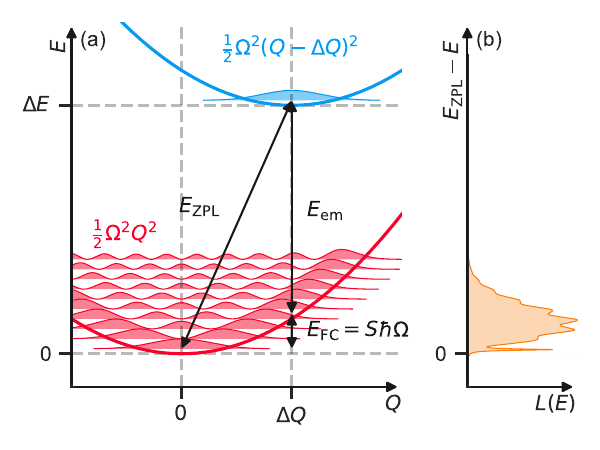}
    \caption{\label{fig:ccd}
        (a) A configuration-coordinate diagram for the model system and (b) a schematic photoluminescence lineshape function $L(E)$.
        Blue indicates the excited state, and red indicates the ground state.
        We assume both states share the same phonon frequency $\Omega$.
        The vibronic wavefunctions are shown schematically.
        The ZPL energy $E_{\rm ZPL}$ and the average emission energy $E_{\rm em}$, as defined in the text, are labeled in black.
        The Franck-Condon energy $E_{\rm FC}$ describes the average energy emitted as phonons and is written in terms of the Huang-Rhys factor $S$.
        We can see that the Huang-Rhys factor is the average number of phonons emitted during the radiative process.
    }
\end{figure}

The effect of this phonon mode on the electronic system, within the single-mode approximation, is shown in the configuration-coordinate diagram in Fig.~\ref{fig:ccd}(a).
One consequence of the coupling between the electronic and vibronic degrees of freedom is that the ground and excited states may not share the same equilibrium geometry.
We define the mass-weighted difference in atomic geometries $\Delta Q$ as
\begin{equation}
    \label{eq:dq}
    {(\Delta Q)}^2 = \sum_I M_I {\lvert {\bf R}_{I,g} - {\bf R}_{I,e} \rvert}^2 \;,
\end{equation}
where $I$ labels the atomic sites, $M_I$ is the $I$th atomic mass, and ${\bf R}_{I,g/e}$ are the coordinates of the $I$th site in the ground ($g$) or excited ($e$) state.

Energy must be conserved during the radiative transition.
When all of the energy is dissipated in the form of a photon (without phonons), the photon is said to be emitted into the ZPL.
$E_{\rm ZPL}$ is the energy of the emitted photons and is given by
\begin{equation}
    \label{eq:ezpl}
    E_{\rm ZPL} = \Delta E + \frac{1}{2}\left(\hbar\Omega_e - \hbar\Omega_g\right) \;,
\end{equation}
where $\Omega_{g/e}$ are the vibrational frequencies of the ground ($g$) and excited ($e$) states.
When $\Omega_{g} = \Omega_{e}$, as we will assume here, $E_{\rm ZPL}$ is identical to the energy separation of the ground and excited states $\Delta E$.

A non-zero $\Delta Q$ means that phonons may be emitted during the radiative emission process.
We assume low temperature, such that only the vibrational ground state is occupied in the initial state of the system.
Since the phonons have energy and energy must be conserved, the emitted photons occur at an energy lower than $E_{\rm ZPL}$.
The vibronic wavefunctions $\chi$ of the ground and excited state are depicted in Fig.~\ref{fig:ccd}(a).
The emission of phonons during the radiative transition gives rise to the phonon sideband observed in luminescence [schematically shown in Fig.~\ref{fig:ccd}(b)].
The average emitted photon energy (including photons emitted into both the ZPL and the phonon sideband) is given by
\begin{align}
    \label{eq:E_em_pre}
    E_{\rm em} &= \sum_{n=0}^{\infty} (\Delta E - n\hbar\Omega) {\lvert \braket{\chi_{e0} \vert \chi_{gn}} \rvert}^2 \\
    \label{eq:E_em}
    &= \Delta E - S\,\hbar\Omega \;.
\end{align}
The second equality in Eq.~(\ref{eq:E_em}) follows from the assumption that the phonon frequencies are the same in the ground and excited state (see Appendix~\ref{app:E_em}).
The dimensionless parameter $S$ is the Huang-Rhys factor, defined as~\cite{alkauskas_first-principles_2014,stoneham_theory_1975}:
\begin{equation}
    \label{eq:hr}
    S = \frac{1}{2\hbar} {(\Delta Q)}^2 \Omega \;.
\end{equation}
The Huang-Rhys factor quantifies the strength of electron-phonon coupling and can be interpreted as the average number of phonons emitted during the radiative emission process.
From Eq.~(\ref{eq:E_em}) we can see that $E_{\rm em}$ must be smaller than or equal to $\Delta E$, as expected.
The total radiative emission rate, including photons emitted into the phonon sideband, is given by (see Appendix~\ref{app:rad}):
\begin{equation}
    \label{eq:rad}
    \Gamma_{\rm R} = {\left( \frac{\mathcal{E}_{\rm eff}}{\mathcal{E}_0} \right)}^2 \frac{n_r \mu^2 (\Delta E)^2 E_{\rm em}}{3 \pi \epsilon_0 c^3 \hbar^4} \;,
\end{equation}
which is reduced from $\Gamma_{\rm R}^{(0)}$ [Eq.~(\ref{eq:rad0}), the rate in the absence of electron-phonon coupling] by a factor of $E_{\rm em} / \Delta E$.

\begin{figure}[!htb]
    \centering
    \includegraphics[width=\columnwidth,height=0.5\textheight,keepaspectratio]{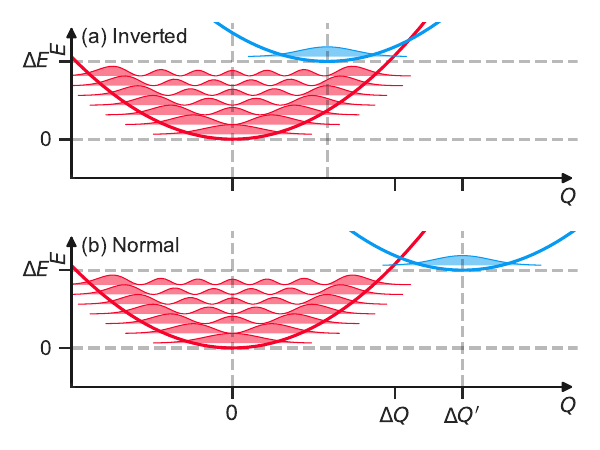}
    \caption{\label{fig:marcus}
        A configuration coordinate diagram in the Marcus (a) inverted and (b) normal regions.
        Blue indicates the excited state, and red indicates the ground state.
        We assume both states share the same phonon frequency $\Omega$.
        The vibronic wavefunctions are shown schematically.
        In the normal region, luminescence is suppressed since the overlap between vibronic wavefunctions of the ground and excited states is negligible.
    }
\end{figure}

$\Gamma_{\rm R}$ is shown in Fig.~\ref{fig:rate} for the case of $S = 3$.
For $\Delta E > 1$~eV, $\Gamma_{\rm R}$ is close to $\Gamma_{\rm R}^{(0)}$, i.e., the impact of phonons on the radiative rate is quite modest.
At lower energies, however, coupling to phonons severely reduces the radiative rate, and at $\Delta E = S \, \hbar \Omega$, $\Gamma_{\rm R}$ drops to zero.
This indicates the transition from the Marcus inverted region ($\Delta E > S \, \hbar\Omega$) to the Marcus region ($\Delta E < S \, \hbar\Omega$)~\cite{stoneham_non-radiative_1981}, see Fig.~\ref{fig:marcus}.
The existence of these two distinct regimes was an important prediction of the seminal theory of Marcus~\cite{marcus_electron_1993}.
In the Marcus region, the overlaps between the vibronic wavefunctions decrease, and luminescence is suppressed (as shown in Fig.~\ref{fig:marcus}).
The strong radiative transitions that we are interested in thus occur all in the Marcus inverted region; this is indeed the most common scenario for transitions at defects in the solid state (as opposed to electron transfer in chemistry, which is what Marcus~\cite{marcus_electron_1993} focused on).

It is common practice to approximate $\Gamma_{\rm R}$ by $\Gamma_{\rm R}^{(0)}$:
as we can see, this approximation is valid for $\Delta E \gg S \, \hbar\Omega$, when the system is well within the Marcus inverted region.
This approximation also benefits from a fortuitous cancellation.
As previously mentioned, local-field effects are commonly ignored and tend to increase the rate, while electron-phonon coupling reduces the rate.
Some cancellation may thus occur between these two effects, improving the agreement between experiment and theory when both are neglected.

Photons that are emitted without emitting phonons correspond to the ZPL.
These are the photons that are in a well-defined quantum state useful for quantum information applications.
The fraction of photons emitted into the ZPL is given by the Debye-Waller factor $e^{-S}$~\cite{stoneham_theory_1975}, and the overall rate of emission into the ZPL is
\begin{equation}
    \label{eq:gam_zpl}
    \Gamma_{\rm ZPL} = e^{-S} \Gamma_{\rm R}^{(0)} \;.
\end{equation}
Equation~(\ref{eq:gam_zpl}) corresponds to the $n=0$ term in Eq.~(\ref{eq:rad_int}).
The Huang-Rhys factor $S$ should therefore be as small as possible to efficiently emit photons in a well-defined quantum state.

\subsubsection{Nonradiative Decay}

\begin{figure}[!htb]
    \centering
    \includegraphics[width=\columnwidth,height=0.5\textheight,keepaspectratio]{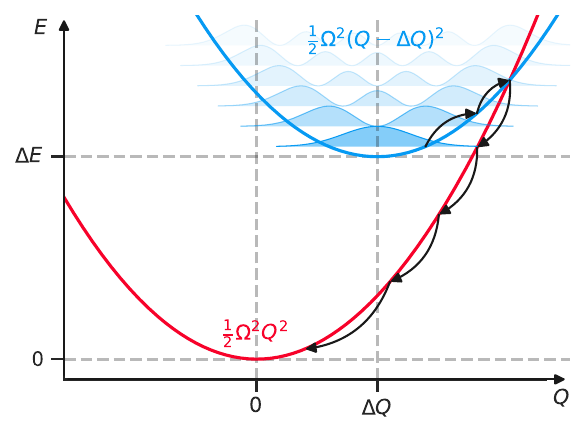}
    \caption{\label{fig:ccd_nr}
        Semiclassical view of the nonradiative process illustrated with a configuration-coordinate diagram for the model system.
        Blue indicates the excited state, and red indicates the ground state.
        We assume both states share the same phonon frequency $\Omega$.
        The vibronic wavefunctions are shown schematically, and the nonradiative process is indicated by the black arrows.
}
\end{figure}

The second consequence of electron-phonon coupling is the introduction of alternative decay mechanisms.
Electron-phonon coupling enables nonradiative relaxation via multiphonon emission~\cite{alkauskas_first-principles_2014,stoneham_theory_1975}.
A semi-classical picture of the multiphonon process is shown in Fig.~\ref{fig:ccd_nr}.
The system is initially in the excited-state configuration with the vibronic states occupied based on thermal equilibrium.
If the system has enough energy, it can surmount the barrier defined by the crossing point between the ground- and excited-state potential energy surfaces.
An electronic transition occurs, and multiple phonons are emitted in the process of relaxing down to the equilibrium configuration of the ground state.

Semiclassically, surmounting the barrier is the rate-limiting step, and this would be a thermally activated process.
However, the actual process is quantum-mechanical, and even at very low temperatures the nonradiative process still occurs due to tunneling through the barrier.
We will show that this nonradiative mechanism can be dominant at small $\Delta E$ even at low temperature.
Going to higher temperatures would increase the nonradiative rate even more; for the purposes of our discussion we focus on the low-temperature case.

Since quantum mechanics governs the behavior of the system the nonradiative transition rate of the multiphonon emission process can be written using Fermi's golden rule~\cite{alkauskas_first-principles_2014,stoneham_theory_1975}:
\begin{multline}
    \label{eq:nr}
    \Gamma_{\rm NR} = \frac{2 \pi}{\hbar} W_{eg}^2 \sum_n {\lvert \braket{\chi_{e0} \lvert \hat{Q} - Q_0 \rvert \chi_{gn}} \rvert}^2 \\
    \times \delta (\Delta E - n \hbar\Omega) \;,
\end{multline}
where $Q_0$ is the geometry for the perturbative expansion of the electron-phonon coupling to linear order.
In this expression we assumed that only the vibrational ground state is occupied in the initial state of the system (as also assumed in Sec.~\ref{sssec:PSB}).
$W_{eg}$ is the electron-phonon coupling matrix element and is system dependent;
we will use a representative value of 0.1~eV/(amu$^{1/2}$~{\AA}), which is comparable to the value for telecom-wavelength emitters in c-BN~\cite{turiansky_telecom-wavelength_2023}, and evaluate the nonradiative transition rate using the implementation in the Nonrad code~\cite{turiansky_nonrad_2021}.

As seen in Fig.~\ref{fig:rate}, the nonradiative transition rate $\Gamma_{\rm NR}$ increases exponentially as $\Delta E$ decreases.
This poses a particular problem for obtaining efficient emission at long wavelengths:
even for a very strong electric-dipole transition ($\mu = 1$~e{\AA}), the nonradiative rate dominates over the radiative rate unless the Huang-Rhys factor is smaller than $S \approx 1$.
Suppressing the nonradiative rate is thus of utmost important for obtaining efficient emitters at energies below 1.5~eV, placing severe constraints on candidate defect centers.

For completeness, we mention that other nonradiative decay mechanisms are possible;
e.g., an Auger-Meitner process could be important in some systems~\cite{abakumov_nonradiative_1991,stoneham_non-radiative_1981,zhao_trap-assisted_2023}.
In the Auger-Meitner process, energy is dissipated by exciting a free carrier to higher energies through the Coulomb interaction.
Since this mechanism is active only in the presence of free carriers, the Auger-Meitner process can be suppressed in samples with low carrier concentrations.
Furthermore, it was found that the process depends weakly on the transition energy~\cite{zhao_trap-assisted_2023}, in contrast to the exponential dependence of the multiphonon process.
Therefore we expect the Auger-Meitner process not to be the dominant decay mechanism at longer wavelengths.

\subsection{Quantum Efficiency}
\label{subsec:qe}
The most common measure of efficiency is the internal quantum efficiency (IQE):
\begin{equation}
    \label{eq:iqe}
    \eta_{\rm IQE} = \frac{\Gamma_{\rm R}}{\Gamma_{\rm R} + \Gamma_{\rm NR}} \;,
\end{equation}
which quantifies the probability of producing a photon (with any energy) per given excitation.
However, as previously noted, only photons in the ZPL are useful for quantum information applications.
A more useful measure is therefore the efficiency of ZPL emission,
\begin{equation}
    \label{eq:qe}
    \eta_{\rm ZPL} = \frac{\Gamma_{\rm ZPL}}{\Gamma_{\rm R} + \Gamma_{\rm NR}} = \frac{e^{-S} \, \Gamma_{\rm R}^{(0)}}{\Gamma_{\rm R} + \Gamma_{\rm NR}} \;.
\end{equation}
The common practice of replacing $\Gamma_{\rm R}$ with $\Gamma_{\rm R}^{(0)}$ means that $\eta_{\rm ZPL} \approx e^{-S} \eta_{\rm IQE}$.
In the literature, it is common to report the Debye-Waller factor and internal quantum efficiency for a given defect.
These values can then be used to estimate $\eta_{\rm ZPL}$, but care should be taken to make sure the limits of the approximation ($S$ is not too large) are not overstepped.

\begin{figure}[!htb]
    \centering
    \includegraphics[width=\columnwidth,height=0.5\textheight,keepaspectratio]{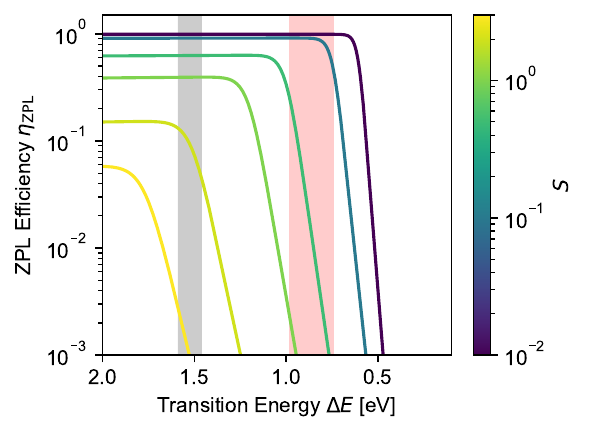}
    \caption{\label{fig:eta_vs_dE}
        Quantum efficiency of ZPL emission $\eta_{\rm ZPL}$ for a strong electric-dipole transition $\mu = 1$~e{\AA}.
        Different colors indicate different values of the Huang-Rhys factor $S$ given by the color bar.
        Energies that fall within the range of telecom wavelengths are shaded in pink, and energies that fall within the free-space communication window are shaded in grey.
    }
\end{figure}

The $\eta_{\rm ZPL}$ as a function of $\Delta E$ for a strong radiative transition ($\mu = 1$~e{\AA}) is shown in Fig.~\ref{fig:eta_vs_dE}.
Larger Huang-Rhys factors clearly suppress $\eta_{\rm ZPL}$, predominantly because of the larger nonradiative recombination rate $\Gamma_{\rm NR}$.
The results confirm the conclusions of Fig.~\ref{fig:rate}: even a relatively small Huang-Rhys factor ($S \approx 1$) will suppress the quantum efficiency below $10^{-3}$ in the telecom-wavelength range.

\begin{figure}[!htb]
    \centering
    \includegraphics[width=\columnwidth,height=0.5\textheight,keepaspectratio]{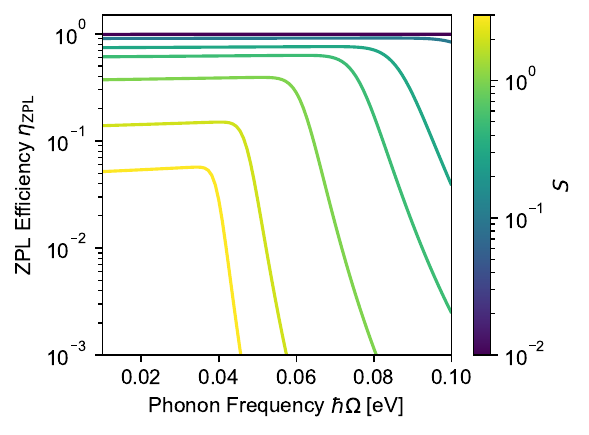}
    \caption{\label{fig:eta_vs_omega}
        The quantum efficiency $\eta_{\rm ZPL}$ as a function of phonon energy $\hbar\Omega$ for $\Delta E = 0.80$~eV.
        Different colors indicate different values of the Huang-Rhys factor $S$ given by the color bar.
    }
\end{figure}

We now discuss approaches to improve the efficiency.
So far, we have kept the phonon energy $\hbar\Omega$ fixed to a value of 100 meV, representative of a case (such as diamond) where high-frequency vibrations dominate the electron-phonon coupling.
In Fig.~\ref{fig:eta_vs_omega}, we investigate the impact of the phonon energy $\hbar\Omega$ on the efficiency $\eta_{\rm ZPL}$, for the case where the emission energy is kept fixed at $\Delta E = 0.80$~eV (in the telecom C-band).
Larger phonon frequencies are clearly detrimental for quantum efficiency.

Phonon frequencies are determined by atomic masses and force constants.
The representative phonon energy is determined by both the host lattice and the defect.
Light host atoms (particularly first-row elements) will inevitably lead to higher frequencies because of their small masses and short bond lengths (which lead to larger force constants).
A host material with heavier atoms would therefore be preferred. 
However, if the defect involves a light impurity, local vibrational modes with larger phonon energies are to be expected;
hydrogen would be particularly problematic in this respect.
We also note that heavier lattices may increase the Huang-Rhys factor, which would negate any improvements in efficiency from lower phonon energies.
This is particularly problematic in soft lattices (e.g., in the halide perovskites~\cite{zhang_defect_2022,zhang_minimizing_2021}).
Still, we suggest that careful exploration of the impact of phonon frequencies is a promising route for improving quantum efficiencies.

\section{Discussion}
\subsection{Tolerating Inefficiency}
Given the inherent difficulty in obtaining a telecom-wavelength transition, or longer wavelengths in general, with high quantum efficiency, a natural question arises:
how much inefficiency can be tolerated?
One method to overcome low quantum efficiency is through coupling to a photonic cavity~\cite{janitz_cavity_2020,parto_cavity-enhanced_2022,peyskens_integration_2019}.

We will consider a cavity with frequency $\omega_c$ tuned to be on resonance with the emitter (i.e., $\hbar\omega_c = E_{\rm ZPL}$).
The strength of coupling between the cavity and emitter is given by~\cite{peyskens_integration_2019}
\begin{equation}
    \label{eq:cav_coupling}
    g = \sqrt{\frac{\xi \, \omega_c}{2 \hbar \epsilon_0 V_c}} \lvert \mu \rvert \cos \theta_d \;,
\end{equation}
where $\theta_d$ is the angle between the polarization of the defect and the cavity.
We will assume perfect alignment and take $\cos\theta_d = 1$.
$\xi = e^{-S}$ accounts for the fact that only emission into the ZPL will couple to the cavity~\cite{kaupp_scaling_2013,parto_cavity-enhanced_2022}.
$V_c$ is the cavity mode volume, which can be designed to give a certain coupling strength.

The cavity is defined by several decay rates.
$\Gamma_p = \omega_c/(2Q)$ is the total decay rate of a cavity with total quality factor $Q$.
$\Gamma_p = \Gamma_c + \kappa$, where $\Gamma_c = \omega_c/(2Q_i)$ is the intrinsic decay rate with intrinsic quality factor $Q_i$.
$\kappa$ is the rate of the decay into the output mode from the cavity, which is the useful light that can be extracted.
For such a cavity, the efficiency of light extraction is given by~\cite{grange_cavity-funneled_2015,peyskens_integration_2019}
\begin{equation}
    \label{eq:eta_cav}
    \eta_{\rm cav} = \frac{\kappa}{(\Gamma_p + \Gamma_e)(1 + \frac{\Gamma_p \Gamma_e}{4g^2})} \;,
\end{equation}
where $\Gamma_e = \Gamma_{\rm R} + \Gamma_{\rm NR}$ is the total decay rate of the emitter.

In the Purcell regime, the cavity enhances the photon density of states leading to an enhancement of the spontaneous emission rate~\cite{burstein_spontaneous_1995}.
To achieve this, the strength of coupling between the cavity and emitter should be weak ($2g < \Gamma_p + \Gamma_e$).
Thus Rabi oscillations between the emitter and cavity are avoided.
Furthermore, the cavity should decay faster than the emitter produces a photon ($\Gamma_p > \Gamma_e$).
This avoids the produced photon from staying in the cavity long enough to interact with emitter again, in other words avoiding high cooperativity~\cite{janitz_cavity_2020}.
The $\kappa$ that optimizes the efficiency of light extraction $\eta_{\rm cav}$ can be obtained by taking a derivative of Eq.~(\ref{eq:eta_cav}) with respect to $\kappa$ (recalling that $\Gamma_p = \Gamma_c + \kappa$) and setting it equal to zero.
The optimal value is given by~\cite{peyskens_integration_2019}
\begin{equation}
    \label{eq:kappa_opt}
    \kappa_{\rm opt} = \Gamma_c \sqrt{ \left(1 + \frac{\Gamma_e}{\Gamma_c}\right) \left(1 + \frac{4g^2}{\Gamma_e\Gamma_c}\right) } \;.
\end{equation}

For a given emitter and the above definitions of the cavity, the remaining parameters to be determined are $V_c$ and $Q_i$.
These parameters are influenced by materials choice, in particular the ability to fabricate photonic structures with that material.
Here we will assume $V_c = 0.1 \lambda_c^3$ where $\lambda_c = c n_r / \omega_c$.
This value should be achievable for emitters operating at visible and longer wavelengths~\cite{robinson_ultrasmall_2005,peyskens_integration_2019,parto_cavity-enhanced_2022}.
We note that constructing photonic cavities is easier at wavelengths longer than visible;
e.g., values as low as $V_c = 10^{-4} \lambda_c^3$ have been demonstrated~\cite{hu_design_2016,choi_self-similar_2017}.
While decreasing $V_c$ increases $\kappa$ and therefore $\eta_{\rm cav}$, it also increases $g$, which could potentially push the system out of the Purcell regime.
Thus the choice of $V_c = 0.1 \lambda_c^3$ is realistic while also avoiding strong coupling.

We then determine $Q_i$ by maximizing $\eta_{\rm cav}$ subject to the constraint that the system is within the Purcell regime.
Furthermore, we restrict $Q_i$ to be less than $10^8$, which is a value used for visible wavelength emitters~\cite{parto_cavity-enhanced_2022}.
While increasing $Q_i$ increases $\eta_{\rm cav}$, the constraint that $\Gamma_p > \Gamma_e$ may no longer be satisfied.
Indeed the main effect of the optimizing $Q_i$ is to maintain $\Gamma_p > \Gamma_e$:
At longer wavelengths, $\Gamma_e$ increases due to the increase in $\Gamma_{\rm NR}$, and therefore, $Q_i$ must decrease to increase $\Gamma_p$ accordingly.

The resulting values of $\eta_{\rm cav}$ are shown in Fig.~\ref{fig:cavity}.
(Optimized values of $Q_i$ and the resulting Purcell enhancement are shown in Appendix~\ref{app:cav}.)
When the transition energy is greater than 1.5~eV, unity efficiency can be obtained even for $S = 3$.
For comparison, without the cavity $\eta_{\rm ZPL}$ was no larger than 6\% when $S = 3$ (yellow line in Fig.~\ref{fig:eta_vs_dE}).
At telecom wavelengths, unity efficiency can be obtained for an emitter with $S < 0.31$, which was only possible for $S < 0.01$ in the absence of a cavity (Fig.~\ref{fig:eta_vs_dE}).

\begin{figure}[!htb]
    \centering
    \includegraphics[width=\columnwidth,height=0.5\textheight,keepaspectratio]{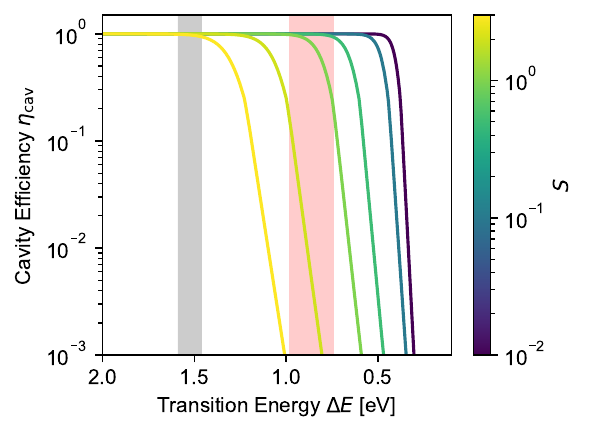}
    \caption{\label{fig:cavity}
        Quantum efficiency of emission from a cavity $\eta_{\rm cav}$ for a strong electric-dipole transition $\mu = 1$~e{\AA}
        and with the cavity parameters specified in the text.
        Different colors indicate different values of the Huang-Rhys factor $S$ given by the color bar.
        Energies that fall within the range of telecom wavelengths are shaded in pink, and energies that fall within the free-space communication window are shaded in grey.
    }
\end{figure}

Our results have shown that obtaining an efficient single-photon emitter is much easier at larger transition energies.
Quantum frequency conversion~\cite{moody_2022_2022,lauk_perspectives_2020} could therefore be used to convert a higher energy photon down to telecom wavelengths for transmission.
The efficiency of conversion can in principle approach unity, and efficiencies exceeding 50\% have been demonstrated~\cite{moody_2022_2022}.
However, the conversion efficiency strongly depends on the specifics of the system, especially with regard to the introduction of noise channels:
conversion from visible or ultraviolet wavelengths to telecom wavelengths is particularly challenging~\cite{moody_2022_2022}.
Quantum frequency conversion from near-infrared (795~nm) to O-band telecom (1342~nm) wavelengths with an efficiency of 33\% is one of the highest values achieved~\cite{yu_entanglement_2020}.

\subsection{Examples}
\label{ssec:examples}

A number of defects that act as single-photon emitters have been observed experimentally; in addition, many have been predicted theoretically.
Some of these defects and their properties, such as emission energy and efficiency, are summarized in Table~\ref{tab:defects}.
From Table~\ref{tab:defects}, we can see that the essential physics captured by our model is confirmed:
smaller transition energies ($\Delta E$) tend to have lower quantum efficiencies $\eta$ due to enhanced nonradiative processes.
As previously mentioned, the NV center and SiV center in diamond have been used for a variety of networking demonstrations~\cite{aharonovich_solid-state_2016,bhaskar_experimental_2020,bernien_heralded_2013}.
$\eta_{\rm ZPL}$ takes a value of 2.3\% for the NV center and 7.5\% for the SiV center, which are relatively high compared to that of the telecom-wavelength emitters discussed below.
The relatively high $\eta_{\rm ZPL}$ values are likely due to the large transition energy $\Delta E$ of 1.95~eV for the NV center and 1.68~eV for the SiV center.

\begin{table*}[htb!]
    \caption{\label{tab:defects}
        Selected host materials and defects that could serve as single-photon emitters.
        Data is shown for NV~\cite{janitz_cavity_2020,gali_ab_2019}, SiV~\cite{janitz_cavity_2020}, and SiV$_2$:H~\cite{mukherjee_telecom_2023} in diamond;
        $V_{\rm B}$-C$_{\rm B}$ and $V_{\rm B}$-Si$_{\rm B}$ in c-BN~\cite{turiansky_telecom-wavelength_2023};
        C$_{\rm B}$-C$_{\rm N}$~\cite{mackoit-sinkeviciene_carbon_2019}, 2-eV emitters~\cite{exarhos_optical_2017,nikolay_direct_2019,li_near-unity_2019}, and the B dangling bond (DB)~\cite{turiansky_dangling_2019,turiansky_impact_2021} in h-BN;
        excitonic emitters in WSe$_2$~\cite{parto_defect_2021,luo_deterministic_2018}; and Er in MgO~\cite{stevenson_erbium-implanted_2022}.
        We assume $\eta_{\rm ZPL} \approx e^{-S} \eta_{\rm IQE}$ here, as described in the text.
        When a range of values is given in the literature, we use the most favorable values to estimate $\eta_{\rm ZPL}$.
        The final column indicates if quantum frequency conversion (QFC) is necessary to transmit the photons over long distances using fiber optics (i.e., the photons need to be converted to telecom wavelengths).
    }
    \begin{ruledtabular}
        \begin{tabular}{c c c c c c c c}
            Host & Defect & $\Delta E$ [eV] & $S$ & $\Gamma_{\rm ZPL}$ [MHz] & $\eta_{\rm IQE}$ & $\eta_{\rm ZPL}$ & Needs QFC? \\
            \hline
            \multirow{3}*{Diamond} & NV$^-$ & 1.95 & 3.5 & 2.1 & 0.76 & $2.3 \times 10^{-2}$ & Yes \\
                             & SiV$^-$ & 1.68 & 0.13--0.29 & 47 & 0.10 & $7.5 \times 10^{-2}$ & Yes \\
                             & (SiV$_2$:H$)^-$\footnote[1]{Tentative attribution from Ref.~\onlinecite{mukherjee_telecom_2023}.} & 1.02 & 0.72 & 3.6 & $2.0 \times 10^{-3}$ & $1.0 \times 10^{-3}$ & No \\
            \hline
            \multirow{2}*{c-BN} & $(V_{\rm B}$-C$_{\rm B})^0$ & 0.95 & 1.5 & 1.2 & $5.4 \times 10^{-4}$ & $1.2 \times 10^{-4}$ & No \\
                                & $(V_{\rm B}$-Si$_{\rm B})^0$ & 0.89 & 1.5 & 1.6 & $3.2 \times 10^{-4}$ & $7.1 \times 10^{-5}$ & No \\
            \hline
            \multirow{3}*{h-BN} & (C$_{\rm B}$-C$_{\rm N})^0$ & 4.31 & 2.0 & 120 & 1.0 & $1.4 \times 10^{-1}$ & Yes \\
                                & ``2-eV emitters'' & 1.6--2.2 & 2--3 & 120 & 0.06--0.87 & $1.2 \times 10^{-1}$ & Yes \\
                                & B DB & 2.06 & 2.3 & 2.6 & 0.06--0.12 & $1.2 \times 10^{-2}$ & Yes \\
            \hline
            WSe$_2$ & Unknown & 1.6--1.7 & $<0.1$~\footnote[2]{No phonon sideband observed in experiments.} & 13 & 0.05 & $5.0 \times 10^{-2}$ & Yes \\
            \hline
            MgO & Er & 0.80 & $<0.1$~\footnotemark[2] & $5.0 \times 10^{-5}$ & 1.0 & $1.0 \times 10^{0}$ & No \\
        \end{tabular}
    \end{ruledtabular}
\end{table*}

The $\eta_{\rm IQE}$ values of the NV and SiV centers are surprisingly low; one might expect that should be closer to unity.
The SiV center has a Huang-Rhys factor $S$ no larger than 0.29, for which even a telecom-wavelength transition should have a high efficiency.
While the NV center has a larger Huang-Rhys factor ($S$=3.5), one would still expect a high $\eta_{\rm IQE}$ within our model given the high transition energy.
Based on our model (Fig.~\ref{fig:eta_vs_dE}) we can conclude that a direct nonradiative process via multiphonon emission is not a limiting factor in the efficiency of the NV and SiV centers.

\begin{figure}[!htb]
    \centering
    \includegraphics[width=\columnwidth,height=0.5\textheight,keepaspectratio]{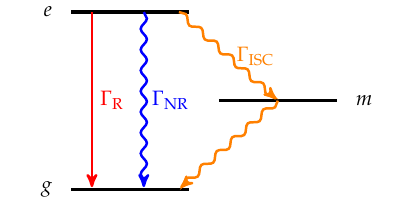}
    \caption{\label{fig:isc}
        Schematic illustration of spin dynamics using an energy level diagram that show the influence of the intersystem crossing on the system.
        The ground $g$ and excited $e$ states are labeled, as well as the metastable excited state $m$ that enables the intersystem crossing.
        Radiative transitions are indicated by solid arrows and nonradiative transitions by wiggly arrows.
        The effect of the metastable state is to draw population away from the optical excited state $e$, thereby lowering the efficiency.
    }
\end{figure}

We suggest that the low $\eta_{\rm IQE}$ values are due to the presence of spin dynamics.
While we do not quantitatively evaluate the rates here, one can qualitatively see from the schematic in Fig.~\ref{fig:isc} that the effect of spin dynamics is to drain away some population from the excited state, leading to a lower overall efficiency.
Indeed, a key component of the NV center---which is integral to its widespread adoption as the prototype quantum defect---is the ability to optically control the ground-state spin, which is enabled by an intersystem crossing between the triplet and singlet manifolds of the center~\cite{gali_ab_2019}.
While spin dynamics can influence the efficiency, it is a second-order concern and can provide an overall benefit in some cases, as exemplified by the NV center.

Similar to spin dynamics, charge dynamics can be understood by including additional states in the model.
While spin dynamics involves transitions within the defect, charge dynamics involves exchanging charge with the local environment (e.g., the bulk bands or nearby traps).
It is generally desirable to avoid charge dynamics completely.
Both charge and spin dynamics can be observed through the photon autocorrelation function, which can be measured spectroscopically~\cite{fishman_photon-emission-correlation_2023,patel_probing_2022}.

We recently predicted $V_{\rm B}$-C$_{\rm B}$ and $V_{\rm B}$-Si$_{\rm B}$ in c-BN as potential NV-center analogues that emit in the telecom range~\cite{turiansky_telecom-wavelength_2023}.
These centers have 
quantum efficiencies of $\sim$10$^{-4}$.
Comparing with a recently observed O-band emitter in diamond~\cite{mukherjee_telecom_2023} is informative.
This emitter has an inferred quantum efficiency of $\sim$10$^{-3}$, and was tentatively attributed to (SiV$_2$:H$)^-$.
The larger quantum efficiency for the diamond emitter could be partly due to the fact that its effective phonon frequency (68.4~meV) is lower that the values for the defects in c-BN ($\sim$100~meV).

In Table~\ref{tab:defects}, there is a notable exception to the low quantum efficiencies observed at telecom wavelengths, namely Er in MgO.
The radiative transition of Er in MgO arises from a transition between $f$ orbitals.
Symmetry dictates that the electric-dipole transition is forbidden, and the emission arises from a magnetic-dipole transition.
No discernible phonon sideband is observed in experiments~\cite{stevenson_erbium-implanted_2022}, indicating that the Er ion is decoupled from the lattice (small $S$).
As a result the coupling to phonons is so weak that the nonradiative process is suppressed, giving high quantum efficiency even though $\Gamma_{\rm ZPL}$ is low.
This result is in line with the predictions of our model if we assume $S \sim 10^{-2}$.
While this might be appealing for having an efficient single-photon emitter in the telecom, there is an obvious drawback:
due to the weakness of the magnetic-dipole transition, the overall rate of single-photon emission is only $\sim$50~Hz~\cite{stevenson_erbium-implanted_2022}.
With cavity coupling, Purcell-enhanced rates up to 21 kHz have been observed~\cite{ourari_indistinguishable_2023}, which are promising but also still orders of magnitude slower than visible and near-IR emitters.

From our survey, some of the most efficient emitters are found in h-BN\@.
The single-photon emission at ultraviolet energies has been attributed to the carbon dimer (C$_{\rm B}$-C$_{\rm N}$)~\cite{mackoit-sinkeviciene_carbon_2019}.
At these energies, the nonradiative process via multiphonon emission is negligible, giving a high internal quantum efficiency.
The overall efficiency of single-photon emission is then governed by the Debye-Waller factor, which is high due to the relatively low Huang-Rhys factor, resulting in a high overall efficiency of emission into the ZPL.

Another class of emitters in h-BN emit in the visible spectrum and are known as the ``2-eV emitters''~\cite{tran_quantum_2016-1,tran_robust_2016,exarhos_optical_2017}.
They are notoriously heterogeneous, as indicated by the spread of values in Table~\ref{tab:defects}.
We proposed the boron dangling bond (DB) as the likely origin of this emission~\cite{turiansky_dangling_2019};
DBs can naturally explain the heterogeneity due to their sensitivity to the local environment~\cite{turiansky_impact_2021}.
Moreover, the spread in observed efficiencies has been linked to the excitation power~\cite{schell_quantum_2018};
the spread can be explained by charge dynamics (photoionization) within the B DB model~\cite{patel_probing_2022}.
The brightest 2-eV emitters are competitive with the carbon dimer in terms of overall efficiency.
However, there is a clear advantage to the 2-eV emitters:
some of the emitters produce photons with a wavelength near 795~nm, which means the already established quantum frequency conversion to telecom wavelengths with an efficiency of 33\%~\cite{yu_entanglement_2020} could be applied to these emitters.
More generally, two-dimensional materials also benefit from improved light-extraction efficiency.

First-principles calculations can be a powerful tool to predict novel defect-based single-photon emitters.
Our work clearly shows the importance of assessing nonradiative decay at longer wavelengths.
Unfortunately, this issue is often overlooked.
In SiC, a near-infrared emitter O$_{\rm C}$-$V_{\rm Si}$ with ZPL energy $\approx$1.2 and Huang-Rhys factor $S = 2.01$ was recently predicted~\cite{kobayashi_oxygen-vacancy_2023}, and also in SiC, the Cl$_{\rm C}$-$V_{\rm Si}$ was predicted to be a telecom-wavelength emitter with Huang-Rhys factor in excess of 3.5~\cite{bulancea-lindvall_chlorine_2023}.
While these defects have lower effective phonon frequencies that are beneficial for efficiency, the low energy of the transition and large Huang-Rhys factors are concerning from the perspective of efficiency.
Given the availability of open-source codes such as Nonrad~\cite{turiansky_nonrad_2021} to evaluate the nonradiative rate from first-principles, we strongly recommend assessing the efficiency of predicted single-photon emitters to determine their feasibility.

\subsection{Future Directions}

In our opinion, the ``Goldilocks'' single-photon emitter has yet to be uncovered.
We are optimistic that efforts to reduce the phonon frequency, as discussed in Sec.~\ref{subsec:qe}, will prove fruitful.
One route of exploration to control the phonon frequency could be to focus on chemical trends.
For example, the group-III nitrides (BN, AlN, and GaN) are being explored as hosts for single-photon emitters~\cite{zhang_material_2020}.
While these materials all contain N, the effective phonon frequency within the one-dimensional model may be lower in GaN than in BN if the phonon mode is dominated by the motion of the heavier Ga atoms.

The observation that transitions between $f$ orbitals leads to small Huang-Rhys factors may also be a fruitful line of exploration.
It may be possible to identify a rare-earth ion and host material combination for which the electric-dipole transition is not forbidden and the radiative rate is sizable.
Along these lines, transition metals may provide a ``middle ground'' that balances moderate emission rates with reduced electron-phonon coupling.

Based on the considerations of cavity coupling, we can suggest that the ``Goldilocks'' emitter may likely have a transmission energy around 1.5~eV.
At this energy, the nonradiative processes are not so severe, and cavity coupling can result in unity efficiency.
Indeed defects whose Huang-Rhys factor is $S$$\approx$3 could still achieve unity efficiency, which is far less restrictive compared to telecom wavelengths.
Moreover, the cavity parameters necessary to realize unity efficiency should be achievable in experiments.
An emitter at 1.5~eV then naturally falls within the window of low atmospheric loss for free-space communication~\cite{kaushal_optical_2017}.
In light of the much higher efficiencies achievable at shorter wavelengths, we suggest that if telecom wavelengths are required for transmission in optical fibers, quantum frequency conversion should be considered alongside direct generation.

\section{Conclusion}

In summary, we have investigated the impact of electron-phonon interactions on the efficiency of defect-based single-photon emitters.
We utilized a model that captures the essential physics of coupling to phonons, based on the formalism of fully first-principles calculations.
Various approximations that have been implicitly used in the literature were discussed, and their validity addressed.
We demonstrated that nonradiative transitions via multiphonon emission become dominant when the transition energy is below 1.5~eV,
dramatically reducing the efficiency with which such defects can produce single photons.
The findings were discussed in the context of values reported in the existing literature.
We proposed engineering approaches to enhance the efficiency, including reducing the phonon frequencies by utilizing chemical trends.
The search is still on to find the optimal single-photon emitter.

\begin{acknowledgments}
    M.E.T. was supported by the U.S. Department of Energy, Office of Science, National Quantum Information Science Research Centers, Co-design Center for Quantum Advantage (C2QA) under contract number DE-SC0012704.
    K.P., G. M., and C.G.VdW. were supported by the National Science Foundation (NSF) through Enabling Quantum Leap: Convergent Accelerated Discovery Foundries for Quantum Materials Science, Engineering and Information (Q-AMASE-i) Award No. DMR-1906325.
    The research used resources of the National Energy Research Scientific Computing Center, a DOE Office of Science User Facility supported by the Office of Science of the U.S. Department of Energy under Contract No. DE-AC02-05CH11231 using NERSC award BES-ERCAP0021021.
\end{acknowledgments}

\section*{Data Availability}
The data that supports the findings of this study can be obtained readily from the figures or are available from the corresponding author upon reasonable request.

\section*{Author Contributions}
Mark E. Turiansky: Writing - original draft (lead); Writing - review \& editing (equal); Formal analysis (lead); Conceptualization (equal); Visualization (lead); Software (lead); Methodology (equal)
Kamyar Parto: Writing - review \& editing (supporting); Conceptualization (supporting).
Galan Moody: Supervision (supporting); Writing - review \& editing (supporting); Conceptualization (supporting); Funding acquisition (supporting).
Chris G. Van de Walle: Supervision (lead); Writing - review \& editing (equal); Conceptualization (equal); Funding acquisition (lead); Methodology (equal).

\appendix

\section{Derivation of the Average Emitted Photon Energy $E_{\rm em}$}
\label{app:E_em}

Here we derive the average emitted photon energy $E_{\rm em}$ including photons emitted into both the phonon sideband and the ZPL Eq.~(\ref{eq:E_em})].
Within the harmonic approximation, the phonon overlap integrals ${\braket{\chi_{e0} \vert \chi_{gn}}}$ can be expressed in terms of the Huang-Rhys factor~\cite{stoneham_theory_1975}:
\begin{equation}
    \label{eq:ovl}
    {\lvert \braket{\chi_{e0} \vert \chi_{gn}} \rvert}^2 = e^{-S} \frac{S^n}{n!} \;.
\end{equation}
Therefore Eq.~(\ref{eq:E_em_pre}) can be written as
\begin{equation}
    \label{eq:E_em_pre_ovl}
    E_{\rm em} = \Delta E \, e^{-S} \sum_{n=0}^{\infty} \frac{S^n}{n!} - \hbar\Omega \, e^{-S} \sum_{n=0}^{\infty} \frac{n S^n}{n!} \;.
\end{equation}
The first summation is just the series expansion of the exponential,
\begin{equation}
    \label{eq:dw_tayseries0}
    e^S = \sum_{n=0}^{\infty} \frac{S^n}{n!} \;.
\end{equation}
Furthermore, we can use
\begin{equation}
    \label{eq:dw_tayseries1}
    S e^{\lambda S} = \frac{d}{d\lambda} e^{\lambda S} = \frac{d}{d\lambda} \sum_{n=0}^{\infty} \frac{(\lambda S)^n}{n!} = \sum_{n=0}^{\infty} \frac{n \lambda^{n-1} S^n}{n!} \;,
\end{equation}
and take $\lambda \rightarrow 1$ to address the second series.
Plugging these results into Eq.~(\ref{eq:E_em_pre_ovl}) gives Eq.~(\ref{eq:E_em}).

\section{Derivation of the Total Radiative Emission Rate $\Gamma_{\rm R}$}
\label{app:rad}

The luminescence intensity $I(\hbar\omega)$ is the number of photons emitted per unit time per unit energy, for a given photon energy $\hbar\omega$~\cite{dreyer_radiative_2020,stoneham_theory_1975}.
Within the Condon approximation and at low temperature where only the ground vibrational level of the excited state is occupied~\cite{markham_electron-nuclear_1965},
\begin{multline}
    \label{eq:I}
    I(\hbar\omega) = {\left( \frac{\mathcal{E}_{\rm eff}}{\mathcal{E}_0} \right)}^2 \frac{n_r \mu^2 (\Delta E)^2}{3 \pi \epsilon_0 c^3 \hbar^4} \\
    \times \sum_{n=0}^{\infty} \hbar\omega \, {\lvert \braket{\chi_{e0} \vert \chi_{gn}} \rvert}^2 \delta(\Delta E - n\hbar\Omega - \hbar\omega) \;.
\end{multline}
The two factors of $\Delta E$ in the numerator come from the fact that the intensity is derived from minimal coupling to the electric field~\cite{dreyer_radiative_2020,stoneham_theory_1975};
thus the coupling entails evaluating momentum matrix elements, which are related to the transition matrix element by $p = i \, m_e (\Delta E / e \hbar) \mu$.
The total radiative rate is an integral of the luminescence intensity over all photon energies:
\begin{align}
    \Gamma_{\rm R} &= \int d(\hbar\omega) \, I(\hbar\omega) \nonumber \\
    \label{eq:rad_int}
    &= \frac{n_r \mu^2 (\Delta E)^2}{3 \pi \epsilon_0 c^3 \hbar^4} \sum_{n=0}^{\infty} (\Delta E - n\hbar\Omega) {\lvert \braket{\chi_{e0} \vert \chi_{gn}} \rvert}^2 \;.
\end{align}
Using the definition of the average emission energy [Eq.~(\ref{eq:E_em_pre})], we arrive at Eq.~(\ref{eq:rad}).

\section{Optimized Cavity Quality Factor}
\label{app:cav}

\begin{figure}[!htb]
    \centering
    \includegraphics[width=\columnwidth,height=0.5\textheight,keepaspectratio]{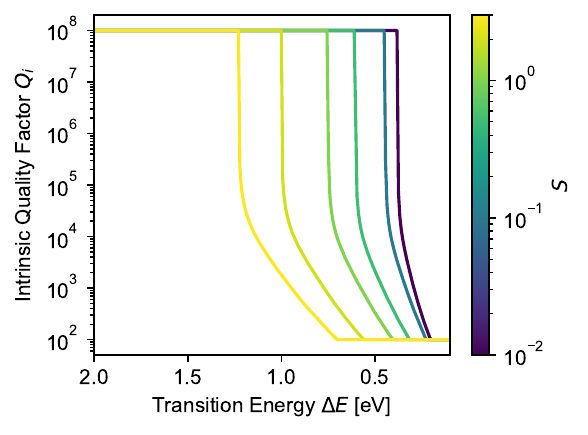}
    \caption{\label{fig:cav_qi}
        The optimized intrinsic quality factor $Q_i$ plotted as a function of the emission energy $\Delta E$ of the single-photon emitter.
        Different colors indicate different values of the Huang-Rhys factor $S$ given by the color bar.
    }
\end{figure}

At a given value of $V_c$, we optimize the intrinsic quality factor $Q_i$ to maximize the efficiency $\eta_{\rm cav}$ while satisfying three constraints.
We constrain the value to be $10^2 < Q_i < 10^8$.
Furthermore, the value must maintain the cavity in the Purcell region, where $2g < \Gamma_p + \Gamma_e$ and $\Gamma_p > \Gamma_e$.
For a given emission energy $\Delta E$ and Huang-Rhys factor $S$, the emitter decay rate $\Gamma_e$ is determined.
For each $\Gamma_e$, we define a fine grid of $Q_i$ values from $10^2$ to $10^8$ and evaluate $\eta_{\rm cav}$ using Eqs.~\ref{eq:eta_cav} and \ref{eq:kappa_opt}.
We then select the value of $Q_i$ for which $\eta_{\rm cav}$ is maximal, while satisfying the constraints.
The resulting values of $Q_i$ are shown in Fig.~\ref{fig:cav_qi}.
When $\Gamma_{\rm NR} < \Gamma_{\rm R}$, the optimal value of $Q_i$ is pinned at the maximal value ($10^8$), but when $\Gamma_{\rm NR} > \Gamma_{\rm R}$, $Q_i$ must drop rapidly to maintain $\Gamma_p > \Gamma_e$.

\begin{figure}[!htb]
    \centering
    \includegraphics[width=\columnwidth,height=0.5\textheight,keepaspectratio]{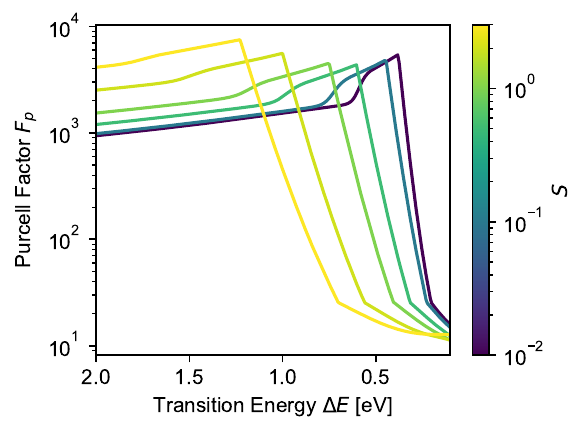}
    \caption{\label{fig:cav_fp}
        The Purcell enhancement $F_p$ resulting from the numerically optimized values of $Q_i$ as a function of the emission energy $\Delta E$ of the single-photon emitter.
        Different colors indicate different values of the Huang-Rhys factor $S$ given by the color bar.
    }
\end{figure}

From the optimized $Q_i$, we can calculate the Purcell enhancement $F_p$ from~\cite{peyskens_integration_2019}
\begin{equation}
    \label{eq:purcell}
    F_p = \frac{3}{4\pi^2} Q \left(\frac{\lambda_c^3}{V_c}\right) \;,
\end{equation}
where we remind the reader that $Q$ here is the total quality factor [i.e., from $\Gamma_p = \omega_c/(2Q) = \Gamma_c + \kappa$, where $\Gamma_c = \omega_c/(2Q_i)$ and $\kappa$ depends on $\Gamma_c$].
Our obtained values of $F_p$ are shown in Fig.~\ref{fig:cav_fp}.
$F_p$ needs to be no larger than $10^4$ to obtain the results shown in the main text.
This is an achievable value given that diamond nanophotonic resonators can theoretically achieve values of $10^5$~\cite{janitz_cavity_2020}.

\bibliographystyle{apsrev4-2}

\end{document}